\documentclass[11pt]{article}
\usepackage{moriond,epsfig}

\def\be{\begin{equation}}
\def\ee{\end{equation}}
\def\bea{\begin{eqnarray}}
\def\eea{\end{eqnarray}}

\begin{document}
\vspace*{4cm}
\title{THE CLUSTER SUBSTRUCTURE - ALIGNMENT CONNECTION}

\author{Manolis Plionis}

\address{Institute of Astronomy \& Astrophysics, National Observatory
of Athens, I.Metaxa \& B.Pavlou, P.Penteli, 152 36 Athens, Greece}

\maketitle\abstracts{
Using the APM cluster data we investigate whether the dynamical status of
clusters is related to the large-scale structure of the Universe. Due
to difficulties in determining unambiguously the dynamical status of
clusters in optical data, our substructure identification method 
has been calibrated using jointly 
ROSAT X-ray and optical data for a subsample of 22 Abell clusters.
We find that cluster substructure is strongly correlated with the
tendency of clusters to be aligned with their nearest neighbour and in
general with the nearby clusters that belong to the same
supercluster. Furthermore, dynamically young clusters are more 
clustered than the overall cluster population.
These are strong indications that cluster develop in a
hierarchical fashion by
anisotropy merging along the large-scale filamentary superclusters within
which they are embedded.}

\section{Overview}
Galaxy clusters occupy a special position in the hierarchy of cosmic 
structure in many respects. Being the largest bound structures in the
universe, they contain hundreds of galaxies and hot X-ray
emitting gas and thus can be detected at large redshifts. Therefore, they
appear to be ideal tools for studying large-scale structure, 
testing theories of structure formation and extracting invaluable 
cosmological information  (cf. B${\rm \ddot{o}}$hringer \cite{bori},
Schindler \cite{Sch00}, Borgani \& Guzzo \cite{borg}).

Below I will review a few issues related to cluster dynamics and the 
cluster large-scale environment. Using the APM cluster catalogue 
(Dalton {\em et al} \cite{Dal97}) and methods calibrated using both optical
and X-ray data I will present significant evidence that the cluster internal
structure and dynamical state is strongly related to their large-scale
distribution.

This contribution is based on different works which
are in preparation with collaborators of different parts of the project
being S.Basilakos, S.Maurocordato, E.Slezak, C.Benoist and others.

\subsection{Cluster Internal Dynamics \& Cosmology}
One of the interesting properties of galaxy clusters is the relation
between their dynamical state and the underlying cosmology. 
Although the physics of cluster formation is complicated 
(cf. Sarazin \cite{Sar01}), it is expected that in an open or a flat with
vacuum-energy contribution universe, clustering effectively freezes 
at high redshifts 
($z\simeq\;\Omega_{m}^{-1}-1$) and thus clusters today should
appear more 
relaxed with weak or no indications of substructure. 
Instead, in a critical density model, such systems continue to form even today
and should appear to be dynamically active (cf. Richstone, Loeb \&
Turner \cite{Ri}, Evrard {\em et al.} \cite{Ev}, Lacey \& Cole \cite{Lac}).
Using the above theoretical expectations as a cosmological tool is
hampered by two facts:
\begin{itemize}
\item[$\odot$] {\it Ambiguity in identifying cluster substructure:}
One has to deal with the issue of unambiguously identifying
cluster substructure, since projection effects in the optical
can conspire to make
cluster images appear having multiple peaks/substructure.
Alot of work has been devoted in attempts to find criteria and methods
to identify cluster substructure (see references in
Kolokotronis {\em et al} \cite{Kol}). 
It is evident from all the available studies
that there is neither agreement on the methods 
utilised nor on the exact frequency of clusters having substructure. 
\item[$\odot$]{\it Unknown physics of cluster merging:}
The clear-cut theoretical expectations regarding the fraction of
clusters expected to be relaxed in different cosmological backgrounds
break-down due to the complicated
physics of cluster merging (cf. Sarazin \cite{Sar01})
and especially due to the uncertainty of the
post-merging relaxation time. In other words, identifying a cluster
with significant substructure does not necessarily mean that this
cluster is dynamically young, but could reflect an ancient merging that
has not relaxed yet to an equilibrium configuration. 
\end{itemize}

Mohr {\em et al.} \cite{Mohr}, Rizza {\em et al.} \cite{Riz} 
and Kolokotronis {\em et al.} \cite{Kol} have investigated 
the frequency of cluster substructure using in a 
complementary fashion optical and X--ray data. The 
advantage of using X--ray data is that the X--ray emission is proportional 
to the square of the gas density (rather than just density in the optical) 
and therefore it is centrally concentrated, a fact which minimises 
projection effects (cf. Sarazin \cite{Sar88}, Schindler \cite{Sch99}). 
The advantage of 
using optical data is the shear size of the available cluster 
catalogues and thus the statistical significance of the emanating results. 
Kolokotronis {\em et al} \cite{Kol} calibrated various substructure measures
using APM and ROSAT data of 22 Abell clusters
and found that in most cases using X--ray or optical data one can
identify substructure unambiguously. Only in $\sim 20\%$ of the
clusters they studied did they find projection effects in
the optical that altered the X-ray definition of substructure. Their
conclusion was that solely optical cluster imaging data can be used 
in order to identify the clusters that have significant substructure.

However, as we discussed previously, it seems that in order to take 
advantage of the different rates of cluster evolution in the different 
cosmological backgrounds one needs {\em (a)} to find criteria of recent
cluster merging and 
{\em (b)} calibrate the results using high-resolution cosmological
hydro simulation, which will provide the expectations of the different
cosmological models.

Such criteria have been born out of numerical simulations
(cf. Roettiger {\em et al.} \cite{Roe93} \cite{Roe99}) 
and are based on the use of 
multiwavlength data, especially optical and X-ray data but radio as
well (cf. Zabuldoff \& Zaritsky \cite{Zab}, Schindler \cite{Sch99},
Sarazin, in this volume).
The criteria are based
on the fact that gas is collisional while galaxies are not and
therefore during the merger of two clumps, containing galaxies and gas,
we expect: ({\it 1}) a difference in the spatial positions of the 
highest peak in the galaxy and gas distribution,
({\it 2}) the X-ray emmiting gas, due to
compression along the merging direction, to be elongated
perpendicularly along this direction and
({\it 3}) temperature gradients to develop due to
the compression and subsequent shock heating of the gas.

Figure 1 presents the smoothed optical (APM) and X-ray (ROSAT pointed 
observations) density 
distributions of 4 Abell clusters, out of which 3 show such
evidence of recent merging, while A2580 seems to be in a relaxed
state (for details of the smoothing procedure see 
Kolokotronis {\em et al} \cite{Kol}). 
Note that BeppoSAX observations of Abell 3266 (de Grandi \&
Molendi \cite{Grandi}) has shown the existence of a temperature gradient
dropping from 10 keV in the cluster core to 5 keV at about 1.5 Mpc distance,
consistent with a merging event, a fact which is also apparent, in
figure 1, from the comparison of the optical and X-ray image of the cluster.
\begin{figure}[t]
\psfig{figure=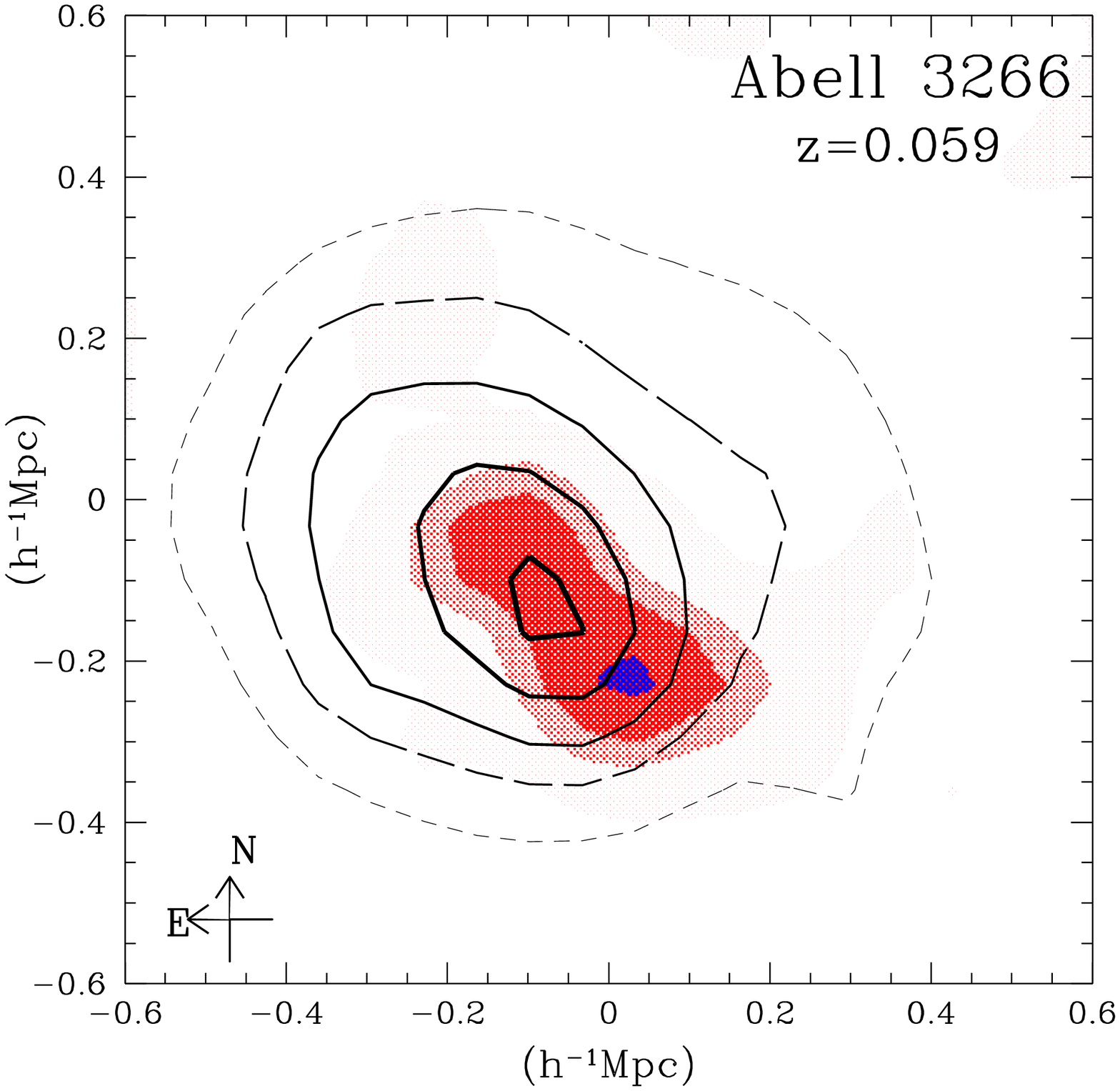,height=2.5in} 
\hfill
\psfig{figure=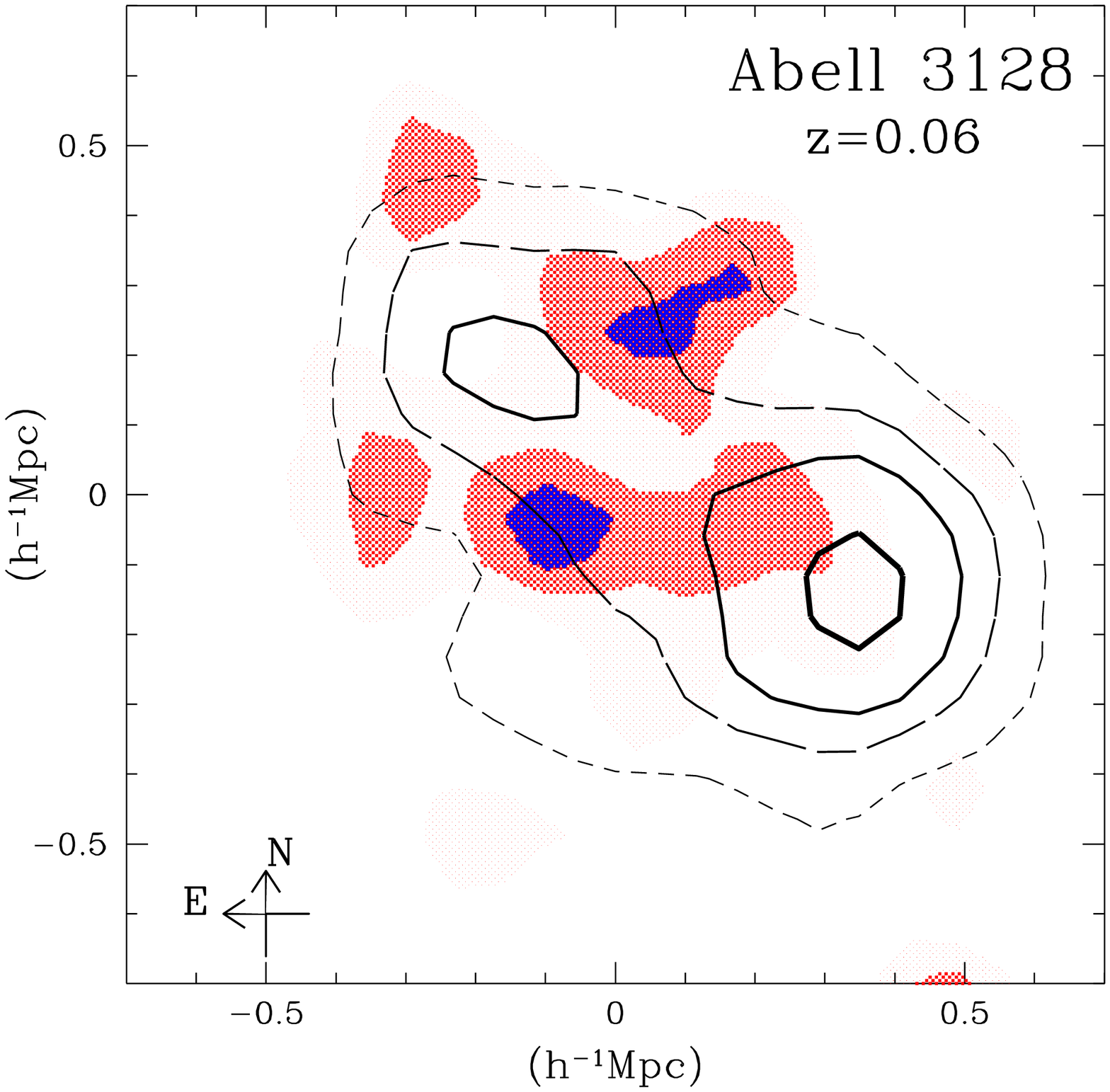,height=2.5in}
\vskip 0.6cm
\psfig{figure=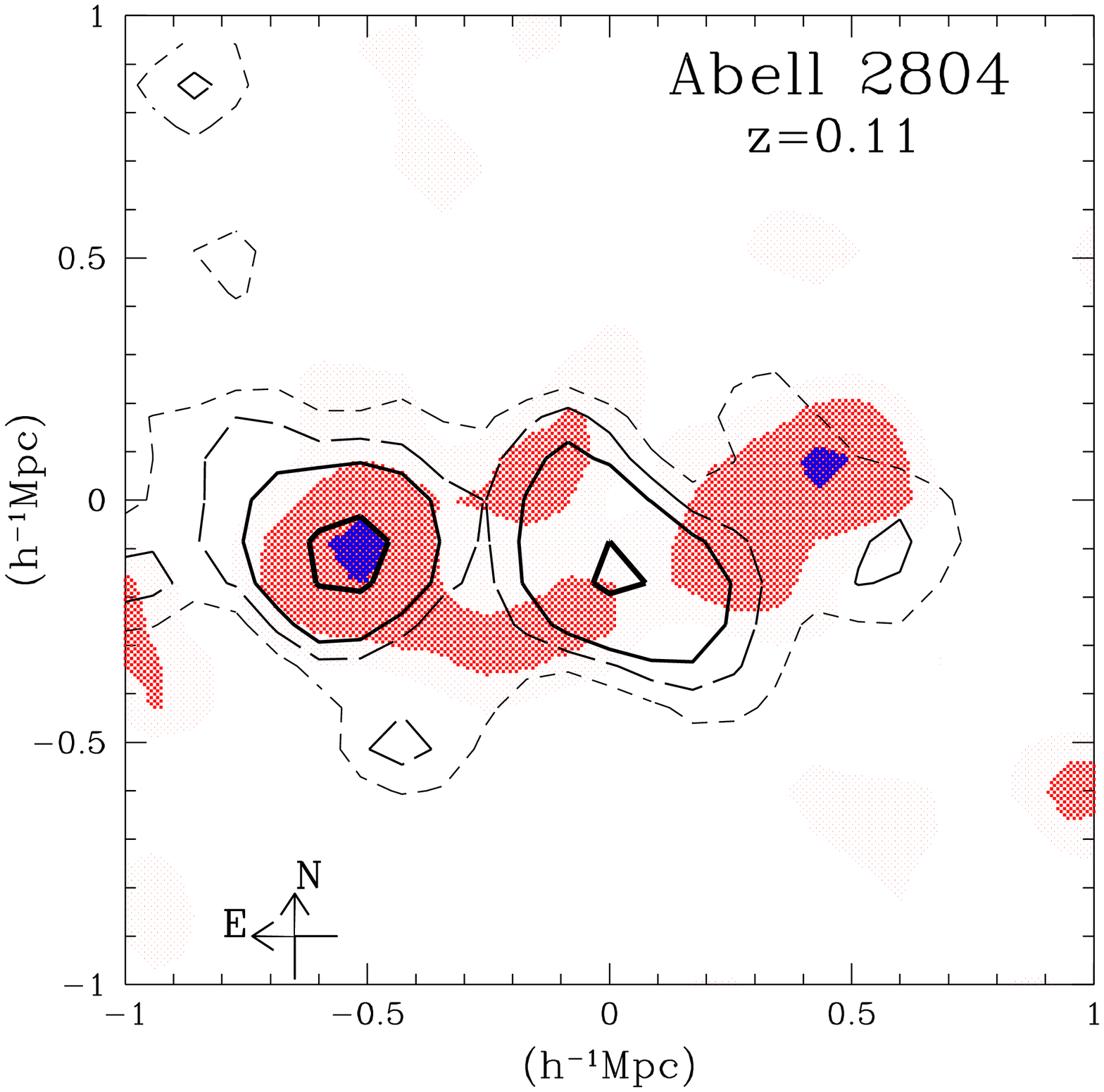,height=2.5in}
\hfill
\psfig{figure=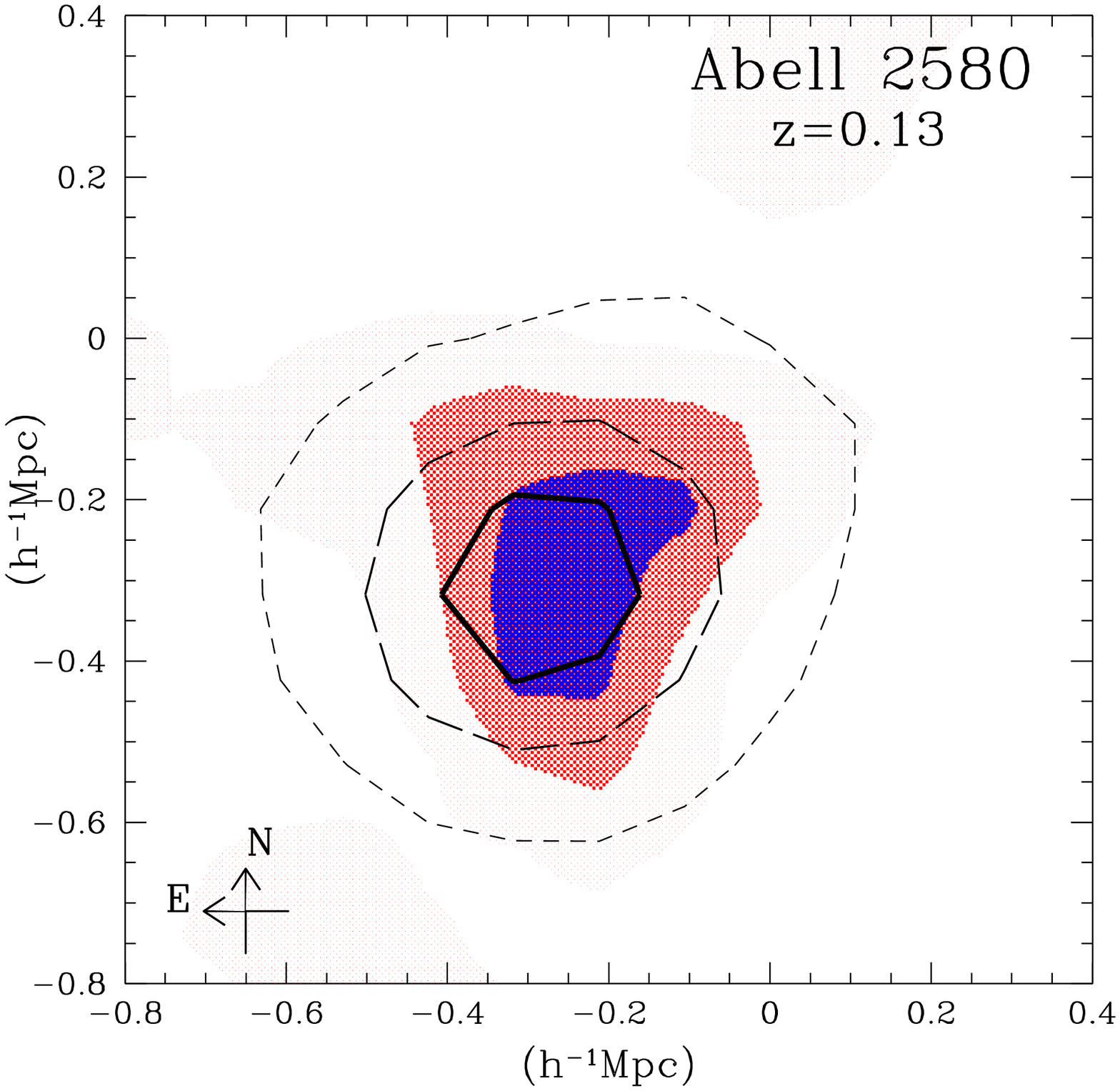,height=2.5in}

\caption{Optical APM (colour) and ROSAT Xray (contour) images
of 4 Abell clusters. Peaks of the APM galaxy distribution is shown 
in blue. Note that appart from A2804, the rest are HRI images.}
\end{figure}

\subsection{Cluster Alignments and Formation Processes}

Another interesting observable, that was thought initially to provide 
strong constraints on theories of galaxy formation, 
is the tendency of clusters to be aligned with their nearest 
neighbour as well as with other clusters that reside in the same 
supercluster (cf. Binggeli \cite{Bin}, West \cite{We89}, 
Plionis \cite{Plio94}). Analytical work of Bond \cite{Bond86} \cite{Bond87}
in which clusters were identified as peaks of an initial Gaussian random 
field, has shown that such alignments, expected naturally to occur in 
"top-down" scenarios, are also found in hierarchical clustering models of 
structure formation like the CDM. These results were 
corroborated with the use of high-resolution N-body simulations by West
{\em et al} \cite{We91}, Splinter {\em et al} \cite{Spli} 
and Onuora \& Thomas \cite{Onu}.
This fact has been explained as the result of an interesting 
property of Gaussian random fields that occurs for 
a wide range of initial conditions and which is the "cross-talk" between 
density fluctuations on different scales. This property is apparently 
also the cause of the observed filamentariness
observed not only in "pancake" models but also in hierarchical models 
of structure formation;
the strength of the effect, however, differs from model to model. 

There is strong evidence that the brightest galaxy (BCGs) in
clusters are aligned
with the orientation of their parent cluster and even with the
orientation of the large-scale filamentary structure within which they
are embedded (cf. Struble \cite{Str90},
West \cite{We94}, Fuller, West \& Bridges \cite{Fu}). 
Furthermore, there is conflicting evidence regarding the
alignment of cluster galaxies in general with the orientation of their
parent cluster (cf. Djorgovski \cite{Dj}, van Kampen \& Rhee \cite{Kamp}, 
Trevese, Cirimele \& Flin \cite{Tre}). It may be that general galaxy
alignments may be present in forming, dynamically young, clusters, while
in relaxed clusters violent and other relaxation
processes may erase such alignment features.
Such seems to be the case of the Abell 85/87/89 complex 
(see Durret {\em et al} \cite{Durr}) and of Abell 521, a
cluster at $z\simeq 0.25$ which is forming at the intersection of two filaments
(Arnaud {\em et al} \cite{Arn}). 
Using wide-field CFHT imaging data of A521, Plionis {\em et al} 
({\em in preparation}) have
found statistical significant alignments not only of the predominantly bright 
but also of fainter
galaxies with the major axis direction of the cluster (figure 2).
\begin{figure}[t]
\psfig{figure=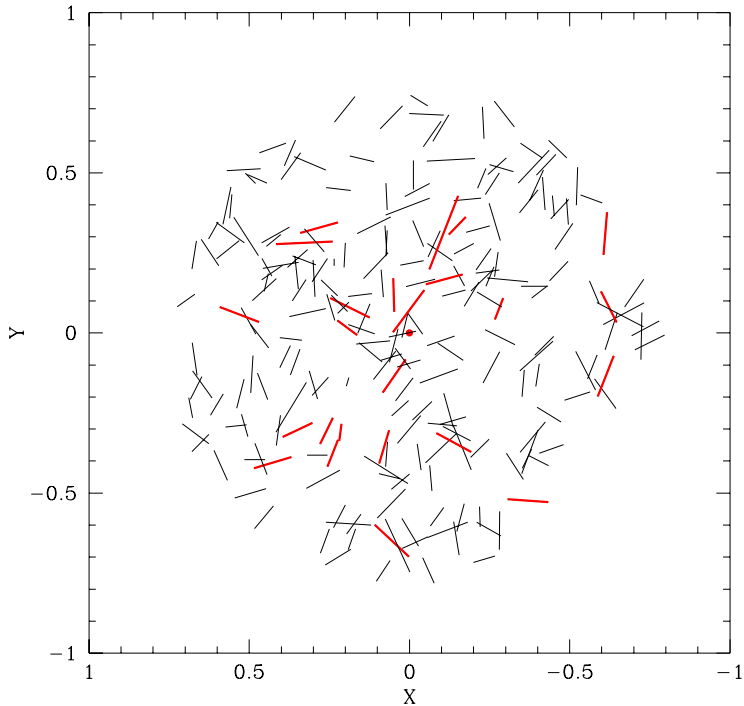,height=4.in}
\hfill \parbox[b]{5.0cm}{\caption{The distribution of galaxy position
angles, within 0.75 $h^{-1}$ Mpc, of A521. Position angles of galaxies 
with $m<m^{*}$ are shown in red. The alignment with the cluster
position angle of $\theta\simeq 140^{\circ}$ is evident.}}
\end{figure}
It is interesting that this direction coincides with 
the direction of the nearest Abell cluster (see figure 3).
\begin{figure}[t]
\psfig{figure=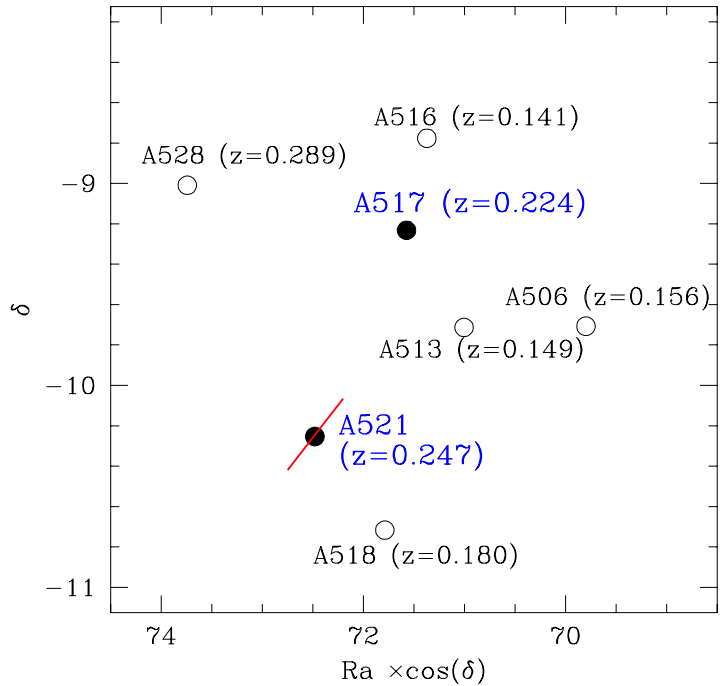,height=3.8in} 
\hfill \parbox[b]{5.0cm}{\caption{The large scale environment
surrounding Abell 521. The major axis direction of A521 is pointing
towards its nearest neighbour A517.}}
\end{figure}
Within the framework of hierarchical clustering, the
anisotropic merger scenario of West \cite{We94}, in which clusters
form by accreting material along the filamentary structure within
which they are embedded, provides an interesting explanation of such
alignments as well as of the observed strong alignment of BCGs 
with their parent cluster orientation.
Evidence supporting this scenario was presented in West, Jones \& 
Forman \cite{We95}
in which they found, using Einstein data, that cluster substructures
seem to be aligned with the orientation of their parent cluster and with the
nearest-neighbouring cluster (see also Novikov {\em et al} \cite{Nov}). 
Such effect has
been observed also in numerical simulations of cluster formation for a
variety of power-spectra (van Haarlem \& van de Weygaert \cite{Haa})

Tidal effects between protoclusters could be an alternative explanation
of galaxy and 
cluster alignments in the popular hierarchical clustering scenario.
However, a back of the envelope calculation shows that in order to
have a fractional effect $f$ on a test particle at a distance of 1
$h^{-1}$ Mpc from a cluster
of mass $M_{cl}$, caused by a perturber at a distance $R$, one needs a perturber
mass of: 
\begin{equation}
M_p \sim f  R^3 M_c
\end{equation}
Therefore for the typical intercluster nearest-neighbour separation
 of $\sim 18$ $h^{-1}$ Mpc, one would get a 5\% effect for 
$M_p \sim 300 M_{cl}$ ! This however
does not exclude the possibility that tides, produced by the cluster itself,
could effect the shape and orientation of cluster members (see
Barnes \& Efstathiou \cite{Barnes} and 
Salvador-Sole \& Solanes \cite{sal} for thorough studies of tidal effects).

\section{Methodology}
Here I briefly present the methods used to determine the dynamical
state of clusters, their shape, orientation and alignment.

\subsection{Substructure Measure - Cluster Centroid Shift}
Evrard {\em et al.} \cite{Ev} and Mohr {\em et al.} \cite{Mohr} 
have suggested as an 
indicator of cluster substructure the shift of the center-of-mass 
position as a function of density threshold above which 
it is estimated. The {\em centroid-shift} ($sc$) is defined
as the distance between the cluster 
center-of-mass, $(x_{\rm o}, y_{\rm o})$, where $x_{\rm o}=\sum\,x_{\rm i}\,
\rho_{\rm i}/\sum\,\rho_{\rm i}$, $y_{\rm o}=\sum\,y_{\rm i}\,\rho_
{\rm i}/\sum\,\rho_{\rm i}$ and the highest cluster density-peak
$(x_{\rm p}, y_{\rm p})$, ie., 
\begin{equation}\label{eq:sc}
sc = \sqrt{(x_{\rm o}-x_{\rm p})^{2}\,+\,(x_{\rm o}-x_{\rm p})^{2}}\,.
\end{equation}
Notice here, that while the cluster center-of-mass changes as a function of 
density threshold, above which we define 
the cluster shape parameters, the
position $(x_{\rm p}, y_{\rm p})$ remains unchanged. According to
Kolokotronis {\em et al}  \cite{Kol}, a large and
significant value of $sc$ is a clear indication of substructure.

The significance of such centroid variations to
the presence of background contamination and random density
fluctuations are quantified using 
Monte Carlo cluster simulations in which, by construction, there is no
substructure.  For each cluster a series of simulated
clusters is produced having the same number of observed galaxies as well as 
a random distribution of background galaxies, determined by the
distance of the cluster and the APM selection function. 
The simulated galaxy distribution follows a King-like profile:

\begin{equation}\label{eq:sb}
\Sigma(r) \propto \left[1\,+\,\left(\frac{r}{r_{\rm c}} \right)^{2}
\right]^{-\alpha} \;,
\end{equation}

\noindent 
where $r_{\rm c}$ is the core radius. We use the weighted, by the
sample size, mean of 
most recent $r_c$ and $\alpha$ determinations 
(cf. Girardi {\em et al.} \cite{Gir95}  \cite{Gir98}), 
i.e., $r_{\rm c} \simeq 0.085 \;h^{-1}$ Mpc and $\alpha \simeq 0.7$. 
We do test the robustness of our results for a plausible range of these
parameters (details can be found in Kolokotronis {\em et al}  \cite{Kol}).
Naturally, we expect the simulated clusters to generate
small $sc$'s and in any case insignificant shifts.
Therefore, for each optical cluster, 1000 such Monte-Carlo clusters are
generated and we derive $\langle sc \rangle_{\rm sim}$
as a function of the same density thresholds
as in the real cluster case. Then, within a search radius of
$0.75 \;h^{-1}$ Mpc from the simulated highest cluster peak,
we calculate the quantity:
\begin{equation}\label{eq:sig}
\sigma =\frac{\langle sc \rangle_{\rm o} - \langle sc 
\rangle_{\rm sim}}{\sigma_{\rm sim}}\;,
\end{equation}
which is a measure of the significance of real centroid shifts
as compared to the simulated, substructure-free clusters. Note that 
$\langle sc \rangle_{\rm o}$
is the average, over three density thresholds, centroid shift
for the real cluster. 

\subsection{Cluster Shape Parameters}

To estimate the cluster parameters we use the familiar moments of 
inertia method (cf. Plionis, Barrow \& Frenk  \cite{Plio91}); with 
$I_{11}=\sum\ w_{i}(r_{i}^{2}-x_{i}^{2})$,
$I_{22}=\sum\ w_{i}(r_{i}^{2}-y_{i}^{2})$,
$I_{12}=I_{21}=-\sum\ w_{i}x_{i}y_{i}$,
where $x_i$ 
and $y_i$ are the Cartesian coordinates of the galaxies and 
$w_i$ is their weight. We, then diagonalize 
the inertia tensor solving the basic equation:
\begin{equation}\label{eq:diag}
det(I_{ij}-\lambda^{2} \; M_{2})=0 ,
\end{equation}
where $M_{2}$ is the $2 \times 2$ unit matrix. The cluster ellipticity
is given by
$\epsilon=1-\frac{\lambda_2}{\lambda_1}$, where $\lambda_i$ are the
 positive eigenvalues with $(\lambda_1>\lambda_2)$.

This method can be applied to the data using either the discrete or smoothed
distribution of galaxies
(for details see Basilakos, Plionis \& Maddox \cite{bas00}):
\begin{itemize}
\item{\em Method 1 ($w_i=1$):} Galaxies within an initial small radius
around the cluster center of mass are used to define 
the initial value of the cluster shape parameters. 
Then, the next nearest galaxy is added to the initial group and the shape 
is recalculated.
Finally we obtain the cluster shape parameters as a function of 
cluster-centric distance.
\item{\em Method 2 ($w_i=\delta\rho/\rho$):} This method is based on 
applying the moment of inertia procedure
on the smoothed galaxy density distribution which we obtain using a Gaussian
 or other kernel on a grid. 
\end{itemize}
{\em Both methods have been found to provide consistent values of the 
cluster orientation but not so of the cluster ellipticity}.

In order to test whether there is any significant bias and 
tendency of the position 
angles to cluster around particular values we estimate their Fourier transform:
\begin{equation}\label{eq:f1}
C_{n} = \left( \frac{2}{N} \right)^{1/2} \sum_{i=1}^{N} \cos 2n\theta_i
\end{equation}
\begin{equation}\label{eq:f2}
S_{n} = \left( \frac{2}{N} \right)^{1/2} \sum_{i=1}^{N} \sin 2n\theta_i
\end{equation}
If the cluster position angles, $\theta_i$, are uniformly distributed 
between $0^{\circ}$ 
and $180^{\circ}$, then both $C_{n}$ and $S_{n}$ have zero mean and unit 
standard deviation. Therefore large values ($>2.5$) indicate significant 
deviation from isotropy.

\subsection{Alignment Measures}
In order to investigate the alignment between cluster orientations,
we define the relative position angle
between cluster pairs by, $\phi_{i,j}\equiv |\theta_i - \theta_j|$. 
In an isotropic distribution we will have
$\langle \phi_{i,j} \rangle \simeq 45^{\circ}$. 
A significant deviation from this would be an indication of an 
anisotropic distribution which can be quantified by 
(Struble \& Peebles \cite{Str95}):
\begin{equation}\label{eq:alin}
\delta=\sum_{i=1}^{N}\frac{\phi_{i,j}}{N}-45
\end{equation}
In an isotropic distribution we have $\langle \delta \rangle \simeq 0$, while
the standard deviation is given by $\sigma=90/\sqrt{12 N}$. 
A significantly negative value of $\delta$ would indicate alignment and 
a positive misalignment.

\section{Results}
\subsection{Cluster substructure}
Applying the above methodology to the $\sim 900$ APM clusters we find that
about 30\% of clusters have significant ($> 3 \sigma$)
substructure. 
\begin{figure}[t]
\psfig{figure=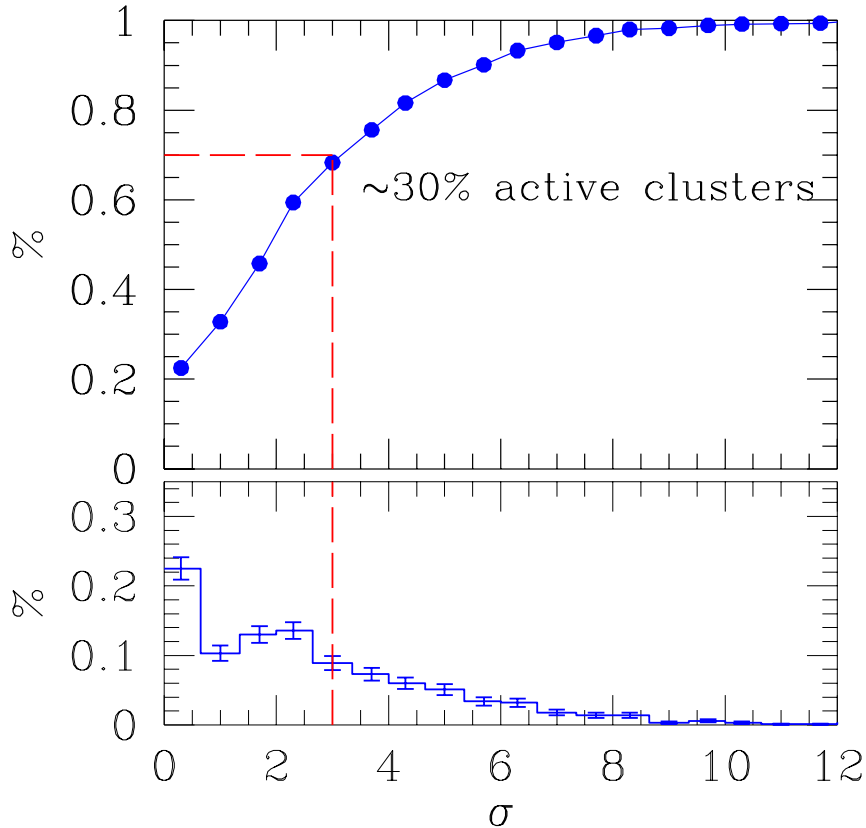,height=5.5in}
\vskip -2.5cm
\caption{The fraction of clusters having
substructure above the indicated significance level.}
\end{figure}
In figure 4 we present the differential and cumulative distribution 
of clusters as a function of substructure significance. Note that
defining as having significant substructure those clusters with $\sigma
>2$ or 2.5 increases the fraction to $\sim$ 50\% and 40\%
respectively. Furthermore, changing the parameters of the Monte-Carlo
clusters changes the actual $\sigma$-values, although their relative
significance rank-order remains unaltered. 
Our results are in accordance with a similar study of the BCS and
REFLEX clusters in which a considerable number of 
clusters do show indications of significant substructures 
(Sch\"{u}ecker {\em et al.}, in preparation).

\subsection{Cluster Alignments}
We have tested whether the well known nearest-neighbour alignment
effect, present in the Abell clusters (cf. Bingelli 1982; Plionis 1994), 
is evident also in the poorer APM clusters.

A necessary prerequisite for this type of analysis is that there is no
orientation bias in the distribution of estimated cluster position angles.
In the lower panel of figure 5 we present the corresponding
distribution for the APM clusters. 
It is evident that the distribution is isotropic, as
it is also quantified by the Fourier analysis. In the upper-panel of
figure 5 we present the distribution of relative position
angles, $\delta\phi$, between APM
nearest-neighbours for two separation limits (one for all separations
and one for separations $< 10 \; h^{-1}$ Mpc). It is
evident that there is significant indication of cluster alignments
in the small separation limit.

\begin{figure}[t]\label{fig:alin}
\psfig{figure=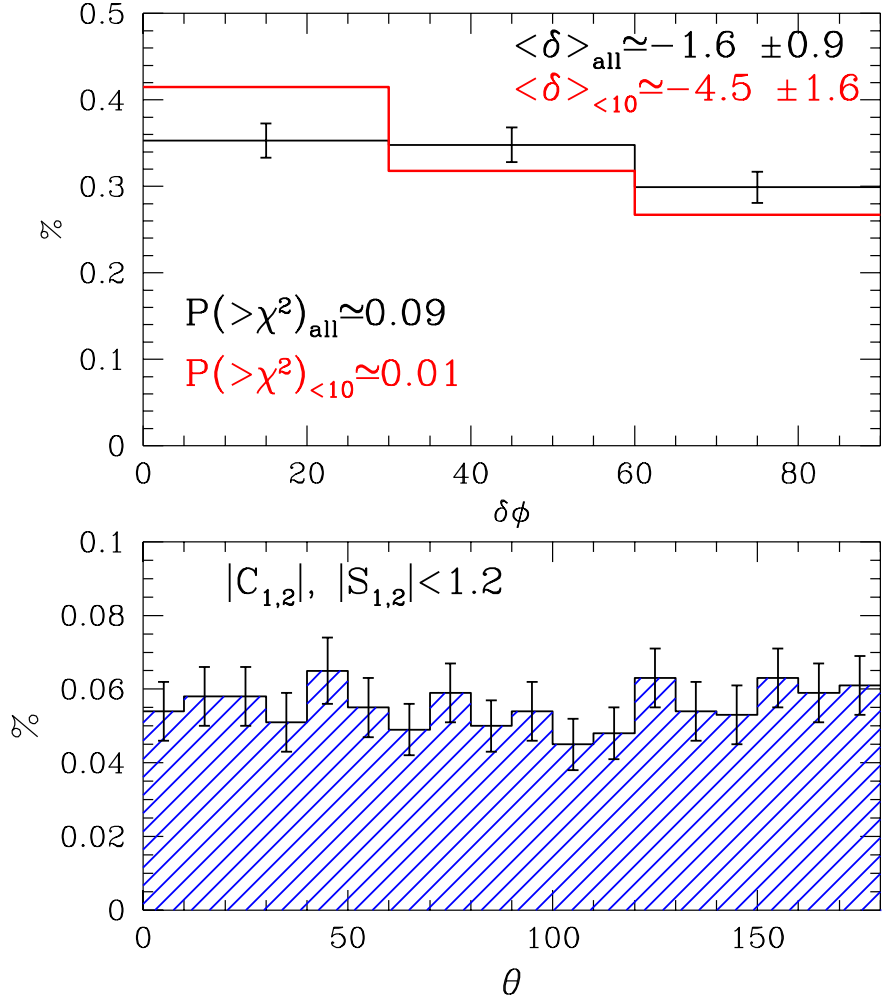,height=5.5in}
\caption{Upper panel: The distribution of relative position angles between
nearest-neighbours. Lower panel: The distribution of APM cluster
position angles.}
\end{figure}

\subsection{Cluster Alignments versus Substructure}
We have correlated the alignment signal with the substructure
significance indication in order to see whether there is any relation
between the large-scale environment, in which the cluster distribution
is embedded, and the internal cluster dynamics. 

In the lower panel of figure 6 we present the alignment signal, 
$\langle \delta \rangle$,
between cluster nearest-neighbours (blue dots) and between all pairs
(open dots) with pair separations $< 20$ $h^{-1}$ Mpc. 
There is a strong correlation
between the strength of the alignment signal and the substructure
significance level. This result, based on the largest cluster sample
available, supports the hierarchical clustering scenario and in
particular the formation of cluster by anisotropic merging along the
filamentary structure within which they are embedded 
(cf. West \cite{We94} \cite{We95}). This is supported also by the fact 
that the mean ellipticity of clusters increases as a function 
substructure significance level, which is in agreement 
with the flatness being a result of accreting
substructures along a preferred direction.
\begin{figure}[t]
\psfig{figure=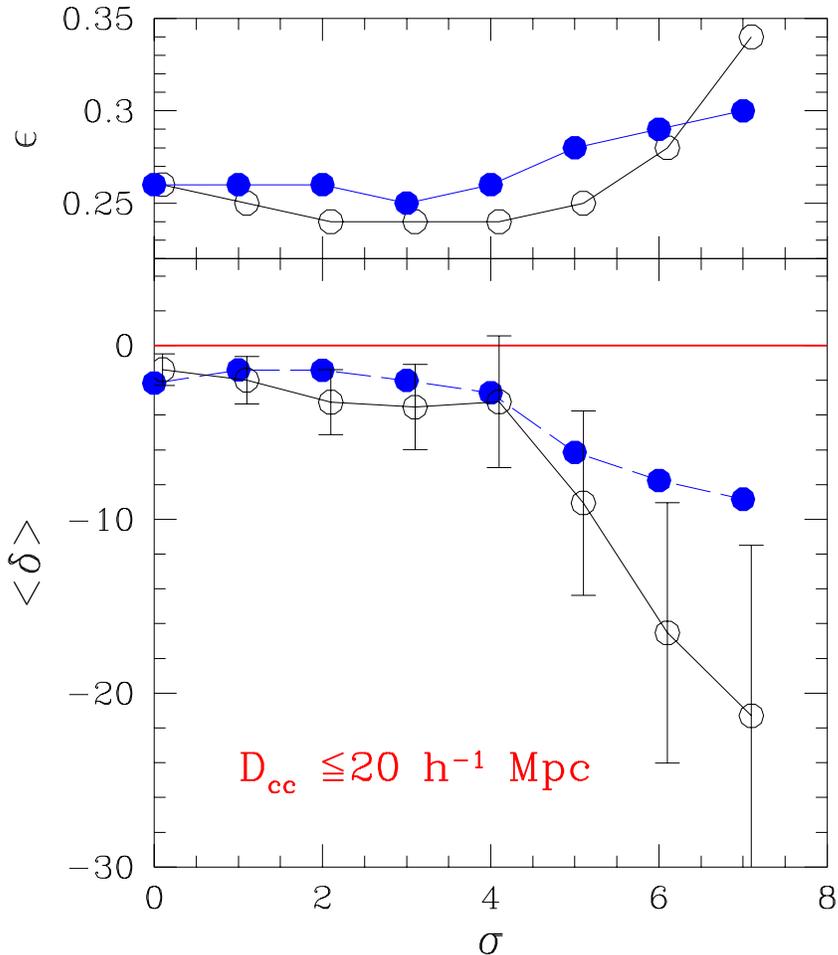,height=5.5in}
\vskip -1cm
\caption{Lower panel: Alignment signal as a function of substructure 
significance level. 
Blue dots correspond to nearest-neighbours and white dots to all
neighbours within 20 $h^{-1}$ Mpc. Upper panel: The mean cluster
ellipticity as a function of $\sigma$.}
\end{figure}

If this view is correct then one would expect that clusters with
significant substructure should be residing preferentially in
high-density environments (superclusters), and this would then
have an imprint in their spatial two-point correlation function. In the
upper panel of figure 7 we present the spatial 2-point
correlation function of all APM clusters (open dots) and of those with
substructure significance $\sigma \ge 4$ (red dots). It is clear that
the latter are significantly more clustered. This can be seen also in
the lower panel were we plot the correlation length, $r_{0}$, as a
function of $\sigma$, which is clearly an increasing function of
cluster substructure significance level. To test whether this effect
could be due to the well-known richness dependence of the correlation strength,
we also present the mean APM richness as a function of $\sigma$ and see
that if any, there is only a small such richness trend. The conclusion
of this correlation function analysis is that indeed the clusters
showing evidence of dynamical activity reside in high-density
environments, as anticipated from the alignment analysis. It is
interesting that such environmental dependence has also been found for
the cooling flow clusters with high mass accretion rates (Loken, Melott
\& Miller \cite{Loken}).
\begin{figure}[t]
\psfig{figure=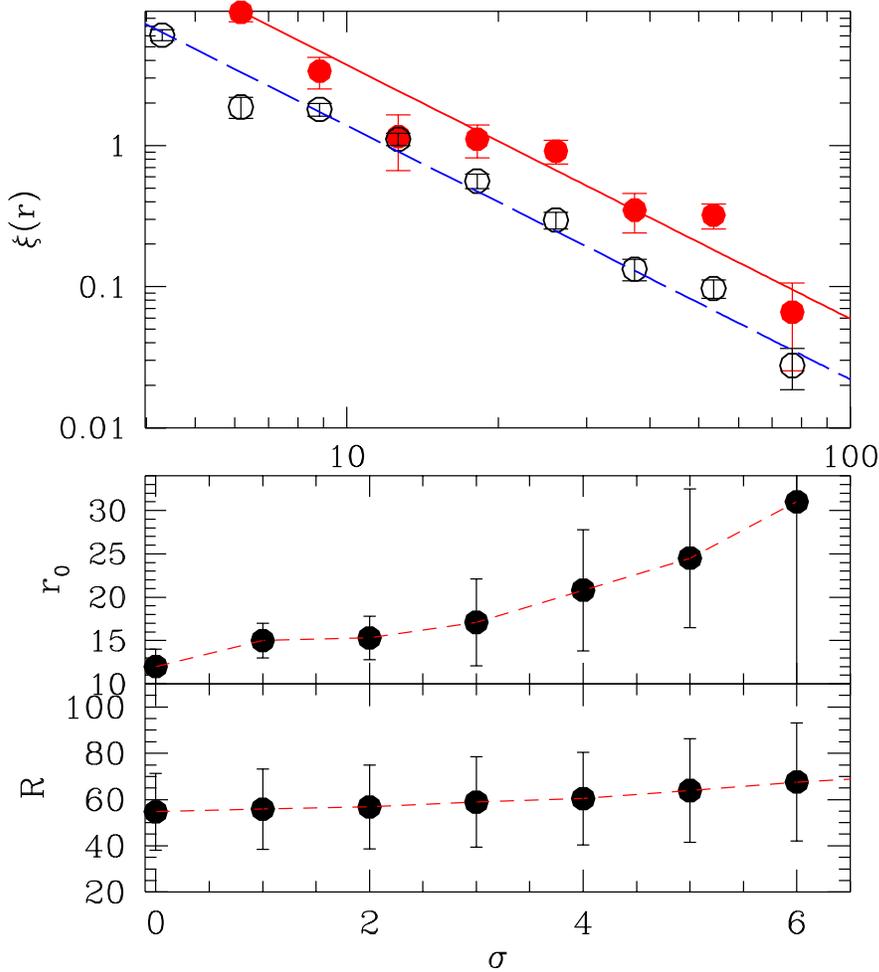,height=5.5in}
\caption{Upper panel: 2-point correlation function of all APM clusters
(open dots) and of $\sigma>4$ clusters. Lower panel: The cluster
correlation length and the mean APM cluster richness
as a function of substructure significance.}
\end{figure}

\subsection{Recent Evolution of Clusters ?}
An interesting exercise is to see whether any of the previously
discussed features show an evolution with redshift, 
since in a low-density (flat or not) 
universe we do not expect to see recent significant 
evolution of the structural or
dynamical parameters of clusters. As a first such test we present in
figure 8 the redshift dependence of the ellipticity for
those clusters that show evidence of significant substructure ($\sigma
\ge 6$). We find correlation coefficient 
$R \sim 0.3$ with probability of no correlation $P<10^{-4}$. A similar
trend is found for the whole APM sample but of lower correlation and
significance. It is interesting that Plionis {\em et al.} \cite{Plio91}
found a similar trend for Abell clusters using the Lick galaxy catalogue.
If this result survives a thorough investigation of possible systematic
effects then it would be interesting to compare this observable
with N-body simulations of different cosmological models in an attempt
to constrain the models.
\begin{figure}[t]
\psfig{figure=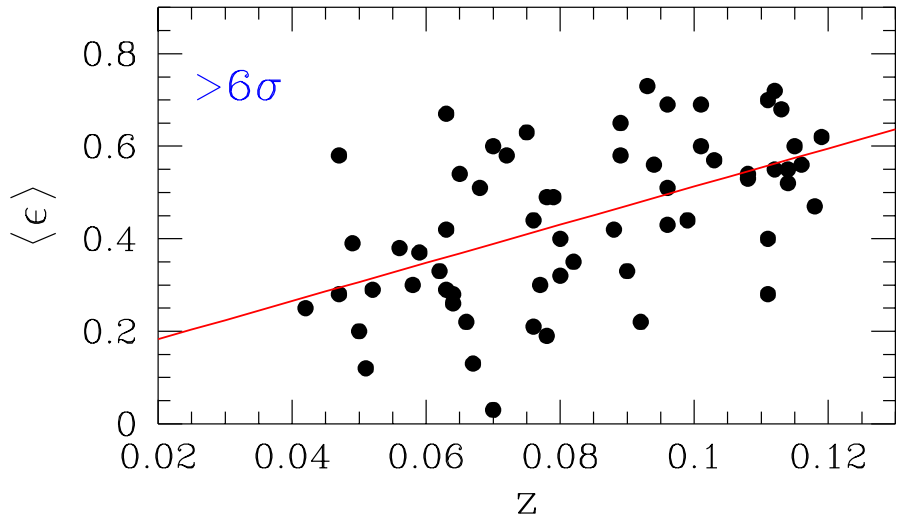,height=5.5in}
\vskip -6.5cm
\caption{Evolution of cluster ellipticity with redshift for those APM
clusters with very significant substructure.}
\end{figure}

\section{Conclusions}
We have presented evidence, based on the large APM cluster sample, 
that there is a strong link between the 
dynamical state of clusters and their large-scale environment.
Clusters showing evidence of
dynamical youth are significantly more aligned with their nearest
neighbours and they are also much more spatially clustered.
This supports
the hierarchical clustering models and in particular the anisotropic
merger scenario of West \cite{We94}.

\section*{Acknowledgments}
I thank all my collaborators for allowing me to present our results
prior to publication.


\section*{References}
\begin{thebibliography}{99}
\bibitem{Arn} Arnaud, M., Maurogordato, S., Slezak, E., Rho, J., 2000,
A\&A, 355, 848
\bibitem{Barnes}Barnes, J., Efstathiou, G., 1987, ApJ, 319, 575
\bibitem{bas00}Basilakos S., Plionis M., Maddox S. J., 2000, MNRAS, 315, 779
\bibitem{Bin}Bingelli B., 1982, AA, 250, 432
\bibitem{Bond86} Bond, J.R., 1986, in {\em Galaxy Distances and
Deviations from the Hubble Flow}, eds. Madore, B.F., Tully, R.B.,
(Dordrecht: Reidel), p.255
\bibitem{Bond87} Bond, J.R., 1987, in {\em Nearly Normal Galaxies},
ed. Faber, S., (New York: Springer-Verlag), p.388
(Dordrecht: Reidel), p.255
\bibitem{borg} Borgani, S. \& Guzzo, L., 2001, Nature, 409, 39
\bibitem{bori}B${\rm \ddot{o}}$hringer H.,
1995, in Proceedings of the 17th Texas Symposium
on Relativistic Astrophysics and Cosmology, eds.
B${\rm \ddot{o}}$hringer H., Tr${\rm\ddot{u}}$mper J., Morfill G. E., 
The New York Academy of Sciences 
\bibitem{Dj}Djorgovski, S., 1983, ApJ, 274, L7
\bibitem{Dal97}Dalton G. B., Maddox S. J., Sutherland W. J., Efstathiou G.,
1997,  MNRAS, 289, 263 
\bibitem{Durr}Durret, F., Forman, W., Gerbal, D., Jones, C., Vikhlinin,
A., 1998, A\&A, 335, 41
\bibitem{Grandi}de Grandi S., Molendi S., 1999, ApJ, 527, L25
\bibitem{Ev}Evrard A.E., Mohr J.J., Fabricant D.G., Geller M.J.,1993, ApJ,
419, L9
\bibitem{Fu}Fuller, T.M., West, M.J. \& Bridges, T.J., 1999, ApJ, 519, 22 
\bibitem{Gir95}Girardi M., Biviano A., Giuricin G., Mardirossian F., 
Mezzetti M., 1995, ApJ, 438, 527
\bibitem{Gir98}Girardi M., Giuricin G., Mardirossian F., Mezzetti M., Boschin 
W., 1998, ApJ, 505, 74
\bibitem{Haa}van Haarlem, M., van de Weygaert, R., 1993, ApJ, 418, 544
\bibitem{Kamp}Kampen van E., Rhee, G.F.R.N., 1990, A\&A, 237, 283
\bibitem{Kol} Kolokotronis, V., Basilakos, S., Plionis, M.,
Georgantopoulos, I., 2001, MNRAS, 320, 49
\bibitem{Lac} Lacey, C., Cole, S., 1996, MNRAS, 262, 627
\bibitem{Loken} Loken, C., Melott, A.L., Miller, C.J., 1999, ApJ, 520, L5
\bibitem{Mohr} Mohr, J.J., Evrard, A.E., Fabricant, D.G.,
Geller, M.J., 1995, ApJ, 447, 8
\bibitem{Nov}Novikov, D. {\em et al.}, 1999, MNRAS, 304, L5
\bibitem{Onu} Onuora, L.I., Thomas, P.A, 2000, MNRAS, 319, 614
\bibitem{Plio91} Plionis, M., Barrow, J.D., Frenk, C.S., 1991, MNRAS,
249, 662
\bibitem{Plio94}Plionis M., 1994, ApJS., 95, 401
\bibitem{Ri}Richstone, D., Loeb, A., Turner, E.L., 1992, ApJ, 393, 477
\bibitem{Riz}Rizza E., Burns J. O., Ledlow M. J., Owen F. N., Voges, W.,
Blito M., 1998, MNRAS, 301, 328
\bibitem{Roe93}Roettiger, K., Burns, J. \& Loken, C., 1993, ApJ, 407,
L53
\bibitem{Roe99}Roettiger, K., Stone, J.M., Burns, J., 1999, ApJ, 518, 594
\bibitem{sal}Salvador-Sole, E. \& Solanes, J.M., 1993, ApJ, 417, 427
\bibitem{Sar88} Sarazin, C.L., 1988, in {\em X-ray Emission from Clusters
of Galaxies}, Cambridge Astrophysics Series, Cambridge Univ. Press.
\bibitem{Sar01} Sarazin, C.L., 2001, in {\em Merging Processes in
clusters of Galaxies}, eds. Feretti, L., Gioia, M., Giovannini, G., 
(Dordrecht: Kluwer).
\bibitem{Sch99} Schindler S., 1999, in Giovanelli F., Sabau-Graziati
L. (eds.), 
proceedings of the Vulcano Workshop 1999, {\em Multifrequency Behaviour
of High Energy Cosmic Sources}, astro-ph/9909042.
\bibitem{Sch00} Schindler S., 2000, in Giovanelli F., G. Mannocchi (eds.),
proceedings of the Vulcano Workshop 2000, {\em Frontier Objects in 
Astrophysics and Particle Physics}, astro-ph/0010319.
\bibitem{Spli}Splinter, R.J., Melott, A.L., Linn, A.M., Buck, C.,
 Tinker, J., 1997, ApJ, 479, 632
\bibitem{Str90}Struble, M.F., 1990, AJ, 99, 743
\bibitem{Str95}Struble, M.F., Peebles, P.J.E., 1985, AJ, 90, 582
\bibitem{Tre}Trevese, D., Cirimele, G., Flin, P., 1992, AJ, 104, 935
\bibitem{We89}West, M. J., 1989, ApJ, 347, 610
\bibitem{We91}West, M. J., Villumsen, J.V., Dekel, A., 1991, ApJ, 369, 287
\bibitem{We94}West, M. J., 1994, MNRAS, 268, 79
\bibitem{We95}West, M. J., Jones C., Forman W., 1995, ApJ, 451, L5
\bibitem{Zab}Zabludoff, A.I. \& Zaritsky, D., 1995, ApJ, 447, L21
  \end{thebibliography}
\end{document}